\documentstyle[12pt]{article}
\begin{document}

\baselineskip 20pt

\newcommand{\sheptitle}
{Heavy Axions from Strong Broken Horizontal Gauge Symmetry}

\newcommand{\shepauthor}
{T. Elliott and S. F. King}

\newcommand{\shepaddress}
{Physics Department\\University of Southampton\\Southampton\\SO9 5NH\\U.K.}

\newcommand{\shepabstract}
{We study the consequences of the existence and breaking of a Peccei--Quinn
symmetry within the context of a dynamical model of electroweak symmetry
breaking based on broken gauged flavour symmetries. We perform an estimate
of the  axion mass by including flavour instanton effects and show that,
for low cut--offs, the axion is sufficiently massive to prevent it from
being phenomenologically unacceptable. We conclude with an examination
of the strong CP problem and show that our axion cannot solve the problem,
though we indicate ways in which the model can be extended so that the strong
CP problem is solved.}

\begin{titlepage}
\hfill SHEP 91/92--22
\vspace{.4in}
\begin{center}
{\Large{\bf \sheptitle}}
\bigskip \\ \shepauthor \\ {\it \shepaddress} \\ \vspace{0.5in}
{\bf Abstract} \bigskip \end{center} \setcounter{page}{0}
\shepabstract
\end{titlepage}

\section{Introduction}

The origin of electroweak symmetry breaking remains an unsolved problem.
While the Higgs sector of the standard model may well be an accurate
reflection of the underlying physical processes driving electroweak symmetry
breaking, and simultaneously giving quarks and leptons their masses, it is
possible that it is only an approximation to some underlying dynamical
mechanism.

The realisation that the top quark, if it exists, must have a large mass
has led many authors to speculate about the role of the top quark in
electroweak symmetry breaking \cite{one}. In particular, it is possible that
a top quark condensate, driven by some as yet undiscovered dynamics, is
responsible for electroweak symmetry breaking. In order to produce a
condensate one needs to induce a 4--quark contact term with a sufficiently
large coupling:
\begin{equation}
{\cal L}={\cal L}_{kin}+G_t({\bar{Q}}_Lt_R)({\bar{t}}_RQ_L),
\end{equation}
where $Q_L^T=(t_L,b_L)$. Recently several authors have begun to consider
the dynamics which may allow such an interaction to appear in the low
energy effective theory \cite{two}.

If we wish to allow the bottom quark to form a condensate, as surely we
must if we wish to maintain maximal correspondence between u,c,t quarks
and d,s,b quarks (and, besides, it will transpire that, in the models we
consider, we cannot give mass to the d,s,b sector any other way), it is
necessary to introduce the additional interaction
\begin{equation}
G_b({\bar{Q}}_Lb_R)({\bar{b}}_RQ_L)
\end{equation}
However, since we shall admit this term, we may include all gauge--invariant
interactions possible including only the top and bottom quarks \cite{three}.
\begin{eqnarray}
\label{tmsm}
{\cal L}={\cal L}_{kin}&+&G_t({\bar{Q}}_Lt_R)({\bar{t}}_RQ_L)+
G_b({\bar{Q}}_Lb_R)({\bar{b}}_RQ_L) \\ \nonumber
&+&G_{tb}\left( ({\bar{Q}}_Lb_R)({\bar{t}}_R^c{\tilde{Q}}_L)+h.c. \right),
\end{eqnarray}
where $\tilde{Q} =i{\sigma}_2Q^c$, and $c$ denotes charge conjugation. If
the contact operators arise from heavy gauge boson exchange \cite{two},
then $G_{tb}\equiv 0$, since the only gauge boson which can mediate
$b_L\rightarrow t_L$ transitions is the $W$. This leads to a problem as
discussed below.

It is easy to show that the Lagrangian density (\ref{tmsm}) is
entirely equivalent to the following, in which the 4--fermi couplings
are reparametrised as Higgs fields without kinetic terms (the kinetic terms
are radiatively induced)
\begin{eqnarray}
{\cal L} = {\cal L}_{kin} &-& {\mu}_t^2{\phi}_t^{\dag}{\phi}_t
+({\bar{Q}}_L{\tilde{\phi}}_tt_R+h.c.) \\ \nonumber
&-& {\mu}_b^2{\phi}_b^{\dag}{\phi}_b
+({\bar{Q}}_L{\phi}_bb_R+h.c.) \\ \nonumber
&-& {\mu}_{tb}^2({\phi}_t^{\dag}{\phi}_b+h.c.)
\end{eqnarray}
If ${\mu}_{tb}\equiv 0$, then the model is the standard two Higgs doublet
model with a $U(1)$ Peccei--Quinn (PQ) symmetry \cite{four}
\begin{equation}
\begin{array}{lll}
Q_L \rightarrow e^{i\alpha}Q_L & t_R \rightarrow e^{-i\alpha}t_R &
b_R \rightarrow e^{-i\alpha}b_R \\ \nonumber
{\phi}_t \rightarrow e^{-2i\alpha}{\phi}_t &
{\phi}_b \rightarrow e^{2i\alpha}{\phi}_b & \nonumber
\end{array}
\end{equation}
Thus, upon spontaneous symmetry breakdown, not only does $SU(2)_L \otimes
U(1)_Y$ get broken to $U(1)_{em}$, but also the ungauged PQ symmetry breaks
resulting in a Goldstone boson. However, the PQ symmetry is explicitly
broken by QCD instanton effects \cite{five}, and the pseudoscalar becomes the
standard
electroweak axion \cite{four,six}. A light pseudoscalar with mass $\sim$ 20 keV
is ruled out experimentally. This is a serious problem for dynamical theories
in which the contact operators are generated by heavy gauge boson exchange.

In what follows we examine a recently proposed model based on broken
gauged chiral flavour symmetries \cite{seven}. It will be seen that this model
also
possesses multiple Higgs doublets and a PQ symmetry. We show,
however, that the PQ symmetry is broken not only by QCD instantons, but also
by instantons associated with the gauged flavour groups and some
technicolour--like
groups. (Henceforth we shall refer to the technicolour--like groups as
ultracolour groups). Since these groups will be strongly coupling at some
scale $\Lambda$,
with $\Lambda \geq 1$ TeV, it will be seen that, under certain circumstances,
the axion will acquire a mass sufficiently large in order to prevent it from
being phenomenologically unacceptable.
\footnote
{Without reference to a particular model, Yamawaki \cite{sevenhalf}
conjectured that instanton effects arising from gauged flavour symmetries
might provided another source of explicit PQ symmetry breaking, therefore
raising the axion mass from its standard QCD value.}

In section 2 we describe the models we shall consider. In section 3 we
calculate estimates of the axion mass in
these models and discuss the implications of its cut--off dependence.
Finally, in section 4, we describe the possible role of this axion in solving
the strong CP problem, and conclude that it does not. We describe extensions
of the models in which the strong CP problem is solved.

\section{``Moose'' Models}

One set of dynamics, which has been proposed recently, is based on a
gauged flavour symmetry which is broken by non--quark condensates
driven by ultracolour interactions \cite{seven}. These condensates in turn
allow the formation of quark condensates resulting in both dynamically
generated quark masses and electroweak symmetry breaking.

In the models which we will consider a multiplicity of both gauged and
ungauged symmetry groups will proliferate. It is therefore convenient,
at this stage, to describe a notation, of which we shall make extensive
use, in which the particle content and particle transformation properties
are completely transparent \cite{eight}. An $SU(n)$ group is represented by the
number
$n$. If the number is circled, then the group is gauged; otherwise it is
a global symmetry. Directed lines correspond to left--handed Weyl fermions.
A line attached to a number, either circled or uncircled, indicates that
the fermions transform under that group. If the line is directed away,
then the fermions transform under the fundamental representation; if
directed towards the number, then the fermions transform under the
conjugate representation.

For simplicity we restrict ourselves at first to an idealised world in
which the down, strange and bottom quarks are absent. We then postulate
the existence of an ultracolour interaction which acts on new
left--handed Weyl fermions, and the gauging of flavour. The new symmetry
groups, where `ga' means gauged and `gl' means global, are
\begin{equation}
\label{groups}
SU(3)_{c} \otimes SU(3)_{f}^{ga} \otimes SU(3)_{L}^{ga} \otimes SU(3)_{R}^{ga}
\otimes SU(3)_{L}^{gl} \otimes SU(3)_{R}^{gl}
\end{equation}
and the quarks $U_L^T\equiv (u_L,c_L,t_L)$, $U_R^{cT}\equiv
(u_R^c,c_R^c,t_R^c)$
and new fermions $a$, $b$, $c$, $d$ transform as
\begin{equation}
\label{transform}
\begin{array}{ccccccccccccccc}
U_{L}     & \sim & (&3&,&1&,&3&,&1&,&1&,&1&) \\
U_{R}^{c} & \sim & (&\overline{3}&,&1&,&1&,&\overline{3}&,&1&,&1&) \\
a         & \sim & (&1&,&3&,&\overline{3}&,&1&,&1&,&1&) \\
b         & \sim & (&1&,&\overline{3}&,&1&,&3&,&1&,&1&) \\
c         & \sim & (&1&,&3&,&1&,&1&,&1&,&\overline{3}&) \\
d         & \sim & (&1&,&\overline{3}&,&1&,&1&,&3&,&1&)
\end{array}
\end{equation}
$SU(3)_{L,R}^{ga}$ are gauged chiral flavour symmetries acting on the left--
and right--handed quarks, respectively. $SU(3)_c$ is the QCD gauge group.
We may introduce a mass matrix connecting the $c$
and $d$ particles. Providing that the elements of the matrix are small
compared to the typical dynamical masses, the explicit breaking of the
global symmetries associated with the $c$ and $d$ particles will be slight.
In figure 1 we give the ``moose'' corresponding to this model. It is
obvious that such a notation enables us to see immediately the structure
of the theory.

We now assume that $SU(3)_f^{ga}$ becomes strongly coupling at some scale
${\Lambda}_f$ and drives the formation of the following condensates
\begin{equation}
\label{condensates}
\begin{array}{c}
<{a_{i}}{d_{i}}> \neq 0 \: , \: <{b_{i}}{c_{i}}> \neq 0 \: , \: i=1,2 \\
\nonumber
<{a_{3}}{b_{3}}> \neq 0 \: , \: <c_3d_3> \neq 0 \nonumber
\end{array}
\end{equation}
Here we do not discuss the precise nature of the vacuum alignment
calculations, but simply refer the interested reader to a recent analysis
\cite{nine}.
It is sufficient to state that if we wish to form a top quark condensate,
then we need a large value of ${\tilde{M}}^U_{33}$, the mass matrix element
connecting the $c_3$ and $d_3$ particles, though not sufficiently large to
break badly the global chiral symmetry. Given such a value, however,
calculations reveal that the condensates of equation (\ref{condensates})
are preferred over
\begin{equation}
<a_id_i> \neq 0 \: , \: <b_ic_i> \neq 0 \: , \: i=1,2,3
\end{equation}

The preferred condensates of (\ref{condensates}) break the groups
$SU(3)_L^{ga}\otimes SU(3)_R^{ga}$ and
$SU(3)_L^{gl}\otimes SU(3)_R^{gl}$ down to the diagonal subgroup
$SU(2)_L\otimes SU(2)_R$ --- the global symmetry of the standard model in
the case of a very massive top quark. Of course, the symmetry will be broken
by the quark mass matrix. That we are able to break the model down to the
standard model together with its associated global symmetries ensures that
we have a G.I.M. mechanism \cite{ten,eight}, the absence of which, hitherto,
has been the
bane of technicolour theories. A G.I.M. mechanism is essential if
${\Lambda}_f$ is to be lowered below about $10^3$ TeV, so that
flavour--changing neutral--currents are suppressed.

The Goldstone bosons generated by the breaking of the gauged
and global groups become the longitudinal components of the gauge
bosons associated with the $SU(3)_L^{ga}$ and $SU(3)_R^{ga}$ groups,
thereby rendering them massive. In the limit $g_{L,R}\rightarrow 0$
(the coupling constants associated with the groups $SU(3)_L^{ga}$ and
$SU(3)_R^{ga}$, respectively) we may calculate the gauge boson mass matrix,
here 16 by 16, which, in general, requires diagonalisation to determine
the mass eigenstates \cite{seven}. Mixing between the L-- and R--sectors can
be so
introduced. It is not necessary here to give the form of the gauge boson
mass matrix. Indeed, it is more edifying, as we shall see, to work with
the gauge eigenstates.

Associated with, for example, the $a$ and $d$ particles there is also
a $U(1)\otimes U(1)$ symmetry, an axial symmetry. The condensates will
break this down to the diagonal subgroup, $U(1)$, generating yet another
Goldstone boson. However the $U(1)$ symmetries are anomalous \cite{eleven} and
thus
the (pseudo--)Goldstone boson receives a mass because of $SU(3)_f$
instanton effects \cite{five}. The dynamics of this mechanism are identical to
those
whereby the $\eta$ acquires a mass $O({\Lambda}_{QCD})$ because of the
explicit breaking of the $U(1)_A$ quark symmetry by QCD instantons. Thus,
our analogue, the ``ultra--$\eta$'' will have a mass $O({\Lambda}_f)$.

The condensates described above thus allow us to induce interactions
between the left-- and right--handed quarks. Working with the gauge boson
gauge eigenstates, the lowest order contribution to a $t_L$--$t_R$
interaction comes from the diagram of figure 2. A cross denotes a
condensate insertion. Of course, there are similar diagrams for
$c_L$--$c_R$ and $u_L$--$u_R$ interactions except that the $(T^8)_{33}$
is replaced by $(T^8)_{11}$ or $(T^8)_{22}$, $T^a$ being the generators
of $SU(3)$. Thus, such diagrams give contributions 4 times smaller than
the $t_L$--$t_R$ interaction.

Such a diagram induces an interaction which, after Fierz re--arrangement,
is of the form
\begin{equation}
g_Lg_RA({\bar{t}}_Lt_R)({\bar{t}}_Rt_L),
\end{equation}
where $A$ is some number with dimensions $[M]^{-2}$ determined by group
theory and the dynamics of the inner loop \cite{seven}. This is precisely the
term
we require in order now to drive the formation of a top quark condensate,
breaking electroweak symmetry and giving the top quark a large dynamical
mass. We must adjust $g_L$ and $g_R$ so that we obtain critical chiral
symmetry breaking. That being so, because the corresponding 4--fermion
interactions of up and charm quarks have an overall coefficient 4 times
smaller than that for the top quark, we will not obtain up and charm quark
condensates in addition.

In fact, to form a top quark condensate the flavour groups must be strongly
coupling, though broken, in addition to the ultracolour groups. Thus, it
is natural to be concerned about the possibility that the strongly coupling
flavour groups will themselves drive other, unwanted, condensates, for
example $<a_iU_i> \neq 0$, which would break the ultracolour groups and
QCD. We simply assume that this does not occur.

Masses for the up and charm quarks are derived from the top mass by
diagrams such as those in figure 3. A blob represents the insertion of a
mass matrix element $({\tilde{M}}^U)_{ab}$ connecting the $c$ and $d$
particles. The corresponding interactions, again after Fierz rearrangements,
are
\begin{equation}
\begin{array}{c}
({\tilde{M}}^U)_{22} B ({\bar{t}}_Lt_R)({\bar{c}}_Rc_L) \\ \nonumber
({\tilde{M}}^U)_{11} B ({\bar{t}}_Lt_R)({\bar{u}}_Ru_L) \nonumber
\end{array}
\end{equation}
If $<{\bar{t}}t> \neq 0$, we see that the up and charm quarks acquire a
mass proportional to (${\tilde{M}}^U)_{ii} <{\bar{t}}t>$.

To extend these ideas to include the down, strange and bottom quarks
is straight--forward. Of course, the left--handed quarks must be placed
into $SU(2)_L$ doublets. Thus, we essentially obtain a moose which
corresponds to 2 copies of the moose of figure 1, except that the
flavour groups acting on the left--handed quarks are compressed into
one flavour group acting on the electroweak doublets, $Q_L$
\begin{equation}
Q_L^T=\left(
\left( \begin{array}{c} u \\ d \end{array} \right)_L,
\left( \begin{array}{c} c \\ s \end{array} \right)_L,
\left( \begin{array}{c} t \\ b \end{array} \right)_L \right).
\end{equation}
Notice that the right--handed quarks $U_R^c$ transform under a different
flavour group from the right--handed quarks $D_R^{cT}=(d_R^c,s_R^c,b_R^c)$.
The moose is in figure 4. (That which has been described above applies
equally to the right--hand side of the moose of figure 4.)

The moose of figure 4, however, has a non--trivial vacuum alignment
calculation associated with it. We do not discuss the problem, but, again,
refer the reader to a recent analysis \cite{nine}.  The result of the analysis,
however,
is that the mass matrices connecting both the c and d and the g and h
particles, from which the quark mass matrices descend, can be simultaneously
diagonalised. Thus, we obtain no understanding of the structure of the
KM mass matrix. We must therefore conclude that the mass matrices themselves
descend from some new dynamics not included in the present moose.

Calculations performed elsewhere reveal that one has to take ${\Lambda}_f$
very large in order to obtain a top quark mass compatible with constraints on
the $\rho$--parameter \cite{seven}. Naively, of course, if a particle obtains
a dynamical
mass from dynamics at the scale $\Lambda$, then the particle should
acquire a mass $O(\Lambda)$. To prevent such an occurrence one has to
fine--tune. We believe, as a general principle, that fine--tuning should
be avoided if possible. Thus, we should like to take ${\Lambda}_f$ $\sim$
1 TeV, this being the lowest scale possible because of phenomenological
constraints. If a quark condensate is responsible for electroweak
symmetry breaking it is not unreasonable to consider the possibility
of a fourth generation of quarks and leptons. We may then take
${\Lambda}_f$ $\sim$ 1 TeV, since there is as yet no phenomenological
limit on the masses of degenerate, fourth generation quarks. Such low
cut--offs, moreover, would enable the models to be falsifiable at, for
example, the S.S.C. We shall call the fourth generation quarks T and B.

The relevant fourth generation moose is readily obtainable from that of
figure 4, simply by replacing the $SU(3)$ flavour groups by $SU(4)$
flavour groups, and the $SU(3)$ global groups by $SU(4)$ global groups.

\section{Calculation of the Axion Mass}

We have seen that interactions of the form $({\bar{Q}}_Lt_R)({\bar{t}}_R
Q_L)$ and $({\bar{Q}}_Lb_R)({\bar{b}}_RQ_L)$ can be obtained from the
moose model. Reflection upon the structure of the moose in figure 4
reveals that we can never induce the term
\begin{equation}
G_{tb}\left( ({\bar{Q}}_Lb_R)({\bar{t}}_R^c{\tilde{Q}}_L)+h.c. \right)
\end{equation}
in our low energy theory. The two operators $({\bar{t}}_Lb_R)({\bar{t}}_R^c
b_L^c)$ and $({\bar{b}}_Lb_R)({\bar{t}}_R^ct_L^c)$ would descend from the
un--Fierzed $({\bar{t}}_L {\gamma}^{\mu} b_L^c)({\bar{t}}_R^c {\gamma}_
{\mu} b_R)$ and $({\bar{b}}_L {\gamma}^{\mu} t_L^c)({\bar{t}}_R^c
{\gamma}_{\mu} b_R)$, and these clearly can never arise from diagrams
of the type in figure 4. Thus $G_{tb} \equiv 0$ and the PQ symmetry is
explicitly broken by instantons only.

Each line (or particle) in the moose of figure 4 transforms under not only
the groups to which the line is attached, but also a $U(1)$ group. We have
already indicated that the non--quark condensates break $U(1)$ symmetries,
but since these symmetries are anomalous, we obtain ``ultra--$\eta$''
particles with large mass. Associated with the three different quark lines
we have the three $U(1)$ symmetries $U(1)_{D_R}\otimes U(1)_Q\otimes
U(1)_{U_R}$; these symmetries can be rewritten in the more familiar form
of $U(1)_B\otimes U(1)_Y\otimes U(1)_{PQ}$. $U(1)_B$ is unbroken by the
quark condensates; $SU(2)_L\otimes U(1)_Y$ is broken down to $U(1)_{em}$,
with a neutral Goldstone boson which becomes the longitudinal component
of $Z^0$. $U(1)_{PQ}$ is ungauged, and its breaking results in a
(pseudo--)Goldstone boson, the axion \cite{four,six}.

We limber up for the calculation of the axion mass in the moose model
by reminding the reader of how to calculate its mass in QCD. This will
enable us to develop further familiarity with the moose notation and
hence allow us to see immediately with which kind of processes flavour
or ultracolour instanton amplitudes are associated. We consider two
flavour QCD, $N_f=2$. The moose appropriate for this case is that shown
in figure 5. The effective QCD instanton vertex is shown in figure 6
and is associated with a process in which $\Delta Q_5=2N_f$, that is,
the axial charge changes by $2N_f$ units, and the vacuum tunnels to a
state of different Pontryagin index \cite{five}.

In order to obtain a non--zero contribution to the vacuum energy density,
one must close off the instanton zero modes. In this way, by coupling
the fermions produced in a tunnelling process to some external, chirality
flipping particles, such as Higgs scalars, one may obtain the diagram of
figure 7. Of course, this diagram explicitly breaks the PQ symmetry,
rendering the axion massive. The amplitude for this process is given by
\cite{five}
\begin{eqnarray}
{\cal M} & = & g_ug_dKe^{-8{\pi}^2/g^2({\Lambda}_{QCD})}
\int \frac{d\rho}{{\rho}^5} {\rho}^{3N_f} (\rho {\Lambda}_{QCD} )^{\beta} \\
\nonumber
& \times & \prod_{i=1,2} \int d^4x \frac{2}{{\pi}^2} {\phi}_i(x)
\frac{1}{[{\rho}^2+x^2]^3},
\end{eqnarray}
where $g_u$ and $g_d$ are the couplings of ${\phi}_1$ to u and
${\phi}_2$ to d, respectively, and $K$  is given by
\begin{equation}
K=\frac{4}{{\pi}^2} \left( \frac{4{\pi}^2}{g^2({\Lambda}_{QCD})} \right)^6
\frac{1}{2} e^{A-B},
\end{equation}
where $A=7.054$, $B=0.360$ and $\beta=11-\frac{2}{3}N_f$. It is to be
understood that there is present a factor $e^{i\theta}$, $\theta$ being the
QCD $\theta$--parameter. Evaluating the integrals (we cut the spatial integrals
off at the lower end with $\rho$, and the instantons with sizes $\sim$
$1/{\Lambda}_{QCD}$ are assumed to dominate), and using Dashen's formula
\cite{twelve}
to give us the mass of a pseudo--Goldstone boson
\begin{equation}
(f_am_a)^2= \: <0\mid [Q^a,[Q^a,{\cal M}]\: ] \mid 0>,
\end{equation}
where $Q^a$ is the charge associated with the axial current, and $f_a$ is
the axion decay constant ($f_a \approx 250$ GeV if the PQ symmetry is
spontaneously broken at the same scale as electroweak symmetry breaking),
we obtain
\begin{equation}
(f_am_a)^2 \sim m_um_d {\Lambda}_{QCD}^2
\end{equation}
Of course, this is precisely the result we expect from dimensional analysis
and the $\eta$ mass.

Comparison of the structure of the QCD moose in figure 5 with the moose
of figure 4, together with knowledge of the effective vertex to which QCD
instantons give rise, allows us to see by inspection the nature of the
processes to which the flavour and ultracolour instantons give rise. It is
easy to see that the flavour and ultracolour groups together provide another
source of PQ symmetry breaking. $SU(3)_U$ instantons force $U_R^c$ and $a$
to rotate together axially; $SU(3)_{fU}$ instantons force $a$ and $b$ to
rotate together, and so on. Thus, from the point of view of explicit PQ
symmetry breaking, the whole moose may be regarded as equivalent to one in
which the quarks transform under two non--abelian groups --- instantons from
both sectors break the PQ symmetry.

Of course, the flavour groups are dynamically broken, and the gauge bosons
become massive. Finite action, non--zero solutions of the Euclidean equations
of motion cease to exist since classical scale invariance is broken, and we
can always rescale non--zero solutions and reduce the action. However, in
Euclidean space, approximate solutions, near which the action varies slowly,
still dominate. The effect of the breaking of classical scale invariance is
to suppress the constrained instanton \cite{thirteen} solutions with a Yukawa
decay at
distances larger than the inverse mass scale. We therefore encode this
behaviour by cutting off the integrals over instanton size at the upper end
by $1/\Lambda$, $\Lambda$ being the scale at which the symmetry is broken.
However, if the gauge groups are strongly coupling at the scale $\Lambda$,
then we may still obtain significant contributions to instanton--induced
amplitudes.

The Higgs particles in our dynamical model are, of course, composite objects,
bound states of quarks. Since left-- and right--handed quarks transform
under different flavour groups, and as there are no current insertions which
can change $t_L$ into $b_{3R}$, there is no one--instanton contribution to
the axion mass. The minimal diagram contains three different instantons,
where, in effect, we have closed off the zero modes of an $SU(3)_Q$
instanton on $SU(3)_U$ and $SU(3)_D$ instantons. However, we now keep the
discussion general and suppose that we have $n$ generations of quarks and
leptons. The diagram is shown in figure 8. The large hatched blobs, of course,
represent the effective vertex of an instanton amplitude; the small unhatched
circles denote a composite Higgs coupling to a fermion; r, g, b distinguish
colours, with tilded colours labelling $SU(3)_{fU}$ colours and doubly
tilded colours labelling $SU(3)_{fD}$ colours. The $e_3$ and $b_3$ particles
are understood to be right--handed, and the $a_3$ and $f_3$ are understood
to be left--handed. The cross, as usual, denotes
a condensate insertion. Inserting condensates is acceptable since we may
suppose that chiral symmetry breaking occurs before confinement, as is
supposed for QCD.

In estimating the contribution of this rather complicated process we assume
that we may perform a naive perturbation theory in the effective instanton
vertices. This naive assumption seems to be vindicated by recent analyses
of the restoration of unitarity in scattering processes by multi--instanton
contributions to the total scattering amplitude \cite{fourteen} and, moreover,
by the
seminal work of Callen, Dashen and Gross \cite{five}. Provided that we work in
the
dilute gas approximation, any ambiguity regarding the zero modes of the
total vacuum tunnelling amplitude is removed, and each sub--process has a
clearly defined number of zero modes.

To calculate the amplitude we need to perform the integral
\begin{equation}
I=\frac{2}{{\pi}^2} \int d^4y \frac{1}{[{\rho}^2_x+(y-x)^2]^\frac{3}{2}}
\frac{1}{[{\rho}^2_z+(y-z)^2]^\frac{3}{2}},
\end{equation}
where we have replaced the Higgs fields by their vacuum expectation values
in anticipation of employing Dashen's formula.
We obtain an angular integral which is a complete elliptic integral of the
second kind. Since we require only an order of magnitude estimate of the
axion mass, we approximate the integrand by its value at $\theta =\pi /2$.
We thus obtain
\begin{equation}
I \approx \frac{4}{[s^2+{\rho}_z^2-{\rho}_x^2]^2} \left(
\sqrt{s^2+{\rho}_z^2}-{\rho}_x \right)^2,
\end{equation}
where $s^2=(z-x)^2$. Recalling that we are working in the dilute gas
approximation, and that the $\rho$ integrals are cut off, $s \gg
max \{ {\rho}_x,{\rho}_z \}$, we may simply take
\begin{equation}
I \approx \frac{4}{s^2}
\end{equation}
and integrate $s$ from $1/{\Lambda}_U$ to $\infty$, ${\Lambda}_U$ being the
scale at which $SU(n)_U$ is broken. Since the groups $SU(3)_U$, $SU(3)_D$ and
$SU(3)_Q$ are all broken at approximately the same scale, we set $\Lambda_U
=\Lambda_D =\Lambda_Q= \Lambda$. We assume, moreover, that the gauge couplings
$g_U$, $g_D$ and $g_Q$ are approximately equal (to g). There are six such legs
emerging from
the $SU(n)_U$ instanton and six from the $SU(n)_D$ instanton, so we need,
\begin{equation}
J = \left[ \int_{1/\Lambda} d^4s \left( \frac{4}{s^2} \right)^6 \right]^2
= 2^{20} {\pi}^4 {\Lambda}^{16}
\end{equation}
For each instanton we must perform an integration over its size, with a
cut--off corresponding to the scale at which instantons become suppressed
because the group is broken. We need
\begin{equation}
{\sigma}_D={\sigma}_U=\int_0^{1/\Lambda} \frac{d\rho}{{\rho}^5}
{\rho}^{3.3} \left( \rho \Lambda \right)^{(\frac{11}{3}n-\frac{2}{3}3)},
\end{equation}
and
\begin{equation}
{\sigma}_Q=\int_0^{1/\Lambda} \frac{d\rho}{{\rho}^5} {\rho}^{3.6}
\left( \rho \Lambda \right)^{(\frac{11}{3}n-\frac{2}{3}6)}.
\end{equation}
So,
\begin{eqnarray}
{\sigma}_U = {\sigma}_D & = & \frac{3}{11n+9} {\Lambda}^{-5} \\
{\sigma}_Q & = & \frac{3}{11n+30} {\Lambda}^{-14}
\end{eqnarray}

This completes all the integrations, so we may put the result together:
\begin{eqnarray}
\label{result}
(f_am_a)^2 & = & K_UK_DK_Q \left( e^{(-8{\pi}^2/g^2)} \right)^3 \\ \nonumber
& \times & (m_U m_D)^3 ({\Lambda}_{fU} {\Lambda}_{fD})^3 \\ \nonumber
& \times & \frac{27}{(11n+9)^2(11n+30)} {\Lambda}^{-8} 2^{20} {\pi}^4,
\end{eqnarray}
where $D$=b or B ($n$=3 or 4), $U$=t or T ($n$=3 or 4), and
\begin{equation}
K_i=\frac{4}{{\pi}^2} \left( \frac{4{\pi}^2}{g^2}
\right)^{2n} \frac{1}{(n-1)!(n-2)!} e^{A+B(n-2-N_f^i)},
\end{equation}
$i=U,D$ or $Q$, and $N_f^U=N_f^D=3$, $N_f^Q=6$. $\Lambda_{fU}$ and
$\Lambda_{fD}$ are taken as the values of the condensate insertions. We
suppose that the scales at which the ultracolour groups break chiral
symmetry are approximately equal, and that $\Lambda_{fU} \approx
\Lambda_{fD} = \Lambda$. Extracting the $n$ dependent
terms from this result, together with the dimensionful quantities, and
calling the remaining constant $C$, we have
\begin{equation}
(f_am_a)^2=C\frac{e^{3nB}}{(11n+9)^2(11n+30)} \frac{1}{[(n-1)!(n-2)!]^3}
\frac{m^3_D m^3_U}{{\Lambda}^2}
\end{equation}

It is well--known that instanton calculations are rather unreliable. Thus
we seek to determine $C$ from general physical principles. The axion
receives a mass from dynamics at the scales associated with the flavour
and ultracolour groups, but arises as a (pseudo--)Goldstone boson
because of spontaneous symmetry breaking at the weak scale. Now, if the
ultracolour groups become strongly coupling at $\sim$ 1 TeV, so that the
associated ultra--$\Pi$ decay constants are $\sim$ 250 GeV --- the
electroweak scale --- we expect to have a theory with no fine--tuning and
thus one in which all particle masses are $\sim$ 1 TeV. Of course, for the
top quark, such a mass is unacceptably large. However, for the T and B
quarks, 1 TeV is perfectly reasonable. The axion will obtain a mass from
dynamics at the scale at which the PQ symmetry is spontaneously broken.
Thus, provided that the couplings of the Higgs particles to the quarks
are of order unity, as they must be since $m_T=g_T<{\phi}_T>$ and
$m_T \approx <{\phi}_T> \approx 1$ TeV, the axion would also have a mass
$\sim$ 1 TeV. (The axion mass in the standard model is so small because
$f_{\pi} \ll f_a$, and the up and down quarks are very light.) We may use
this general argument to determine C. Taking $f_a=250$ GeV, all mass
scales $\sim$ 1 TeV, and $n=4$ (4 generations) we obtain
\begin{equation}
C=3.0 \times 10^5
\end{equation}
This value of $C$ corresponds to taking ${\alpha}_Q={\alpha}_U=
{\alpha}_D\approx 7$ at a scale of 1 TeV. This agrees with our expectation
that the flavour groups need to be strongly coupling in order to drive
quark condensates. Calculations performed elsewhere \cite{eight} indicate that
$\alpha \approx 30$ is required to obtain quark condensates. However,
these calculations are based on a truncation of a Dyson--Schwinger
equation and thus should be regarded as indicating the general qualitative
features of the theory, rather than providing reliable estimates of
physically interesting quantities.

For $n=3$ and $n=4$, we have
\begin{eqnarray}
\label{threefamilies}
f_am_a & = & \frac{2.9}{\Lambda} \sqrt{m_t^3m_b^3}  \; \; \; \; n=3 \\
f_am_a & = & \frac{1}{4\Lambda}  \sqrt{m_T^3m_B^3}  \; \; \; \; n=4
\end{eqnarray}
Taking $m_t=125$ GeV and $m_b=5$ GeV, for
$\Lambda =1$ TeV, we obtain $m_a=183$ MeV for 3 generations of quarks,
and for $m_T=m_B=1$ TeV, $m_a=1$ TeV for 4 generations of quarks, as
arranged.

Clearly $m_a=1$ TeV is acceptable, and possibly $m_a=183$ MeV is also
acceptable phenomenologically \cite{fifteen}. In the above estimates we
have assumed $\Lambda =1$ TeV, and for the three generation model we have
taken $m_t=125$ GeV rather than the much larger value expected from
standard calculations \cite{one}. If these standard calculations are at all
reliable, then we would expect to have a much larger value of $\Lambda$
in the three generation case, for a reasonable value of $m_t$. For example,
$m_t \approx 230$ GeV requires $\Lambda \approx 10^{15}$ GeV \cite{one}.
For such a large value of $\Lambda$ it is clear from equation
(\ref{threefamilies}) that the axion receives a contribution to its mass from
flavour instantons of $f_am_a \sim 10^{-10}$ $(GeV)^2$, which is much
too small.  Thus, our calculations rule out the possibility of three families
\footnote
{Interestingly there is another problem with the three family model due to
the decay $\Upsilon
\rightarrow \gamma H_0'$, where $H_0'$ is a Higgs scalar associated with
the $<\bar{b}b>$ condensates. In the three generation model with
$G_{tb}=0$ we expect $m_{H_0'}<2m_b$, which is apparently ruled out by the
above process.}
with a top quark condensate induced by heavy gauge boson exchange, due to
an unacceptably light axion.

\section{The Strong CP Problem}

The standard model possesses two possible sources of strong CP violation:
the QCD $\theta$--parameter; phases in the quark mass matrix. In the
absence of colour instantons one may re--define the quark fields so that
the mass matrix is real, and in the absence of a quark mass matrix one may
re--define the quark fields and set $\theta =0$. However, in the presence
of instantons and phases in the quark mass matrix, there is a potential
conflict between the direction chosen for zero CP violation on the vacuum
manifold. This conflict would result in overall strong CP violation.
Experimentally
it is known that $\theta + arg \: det \: M = \bar{\theta} \leq 10^{-9}$.
This is the strong CP problem. For a review of the strong CP problem see,
for example, \cite{fifteen} and references therein.

In the two Higgs doublet extended standard model, the
Peccei--Quinn--Wilczek--Weinberg
axion provides a natural understanding of the absence of strong CP violation.
Through its coupling to the anomaly as $\frac{a}{32{\pi}^2f_a}F^a_{\mu \nu}
{\tilde{F}}^a_{\mu \nu}$, and after re--defining the axion field by
$a \rightarrow a - f_a\bar{\theta}$, one can show that the vacuum of
smallest energy density corresponds to $<a>=0$. Thus, the
axion forces the two directions of zero CP violation on the vacuum manifold
to be consistent. Unfortunately there is no empirical evidence to suggest
that the PQWW axion exists.

Given the occurrence of multiple Higgs doublets in the present model, and
a massive axion, it is natural to ask whether the strong CP problem is
solved here. We begin by demonstrating that there is only one overall
$\theta$--parameter in the moose of figure 4, rather than 6.

We may work at a sufficiently high energy scale (temperature) above which
the vacuum has undergone a phase transition and entered a phase in which
all the condensates have evaporated. In this phase, all particles are massless,
except that the c--d and g--h mass matrices remain because they are
non--dynamical. Because all dynamical masses vanish, we are at liberty to
perform axial re-definitions of the particle fields. Consider, for example,
the $U_R^c$ line which has associated with it two $\theta$--parameters,
$\theta_c$, the colour $\theta$--parameter, and $\theta_U$, the flavour
$\theta$--parameter of the right--handed u,c,t (,T) quarks. Now we
perform an axial re-definition of the $U_R^c$ fields. Because the
functional measure is not invariant under such a re--definition \cite{sixteen},
both $\theta_c$ and $\theta_U$ change. If we change the phase of the
$U_R^c$ fields by $\alpha$, then $\theta_c \rightarrow \theta_c -2n\alpha$
and $\theta_U \rightarrow \theta_U +2n_c\alpha$ ($n$ is the number of
generations, $n_c$ the number of quark colours). Setting
$2n\alpha =\theta_c$ we may rotate $\theta_c$ to zero at the expense of
changing $\theta_U$ to $\theta_U+\frac{n_c}{n}\theta_c$. By rotating other
fields we may move all $\theta$--parameters onto, for example, the $SU(3)_{fU}$
group. All other $\theta$--parameters would be zero and thus would give no
CP violation. The $\theta$--parameter associated with the $SU(3)_{fU}$
group would be $\theta_{fU}'=\theta_{fU}+\theta_{fD}+\theta_c+\frac{n}{n_c}
(\theta_Q+\theta_U+ \theta_D)$, in an obvious notation. Moreover, by performing
axial re--definitions of the c and d fields, we may rotate $\theta_{fU}'$
to zero and place it, instead, as a phase in the c--d mass matrix. Any
other configurations of $\theta$--parameters, experimentally, must be
identical to that which we have just obtained.

Our axion is a composite of quarks, the quarks feeling only colour and
flavour forces. Therefore the axion does not couple to ultracolour
fields. In particular, the axion cannot be used to argue that the state
of minimum vacuum energy density occurs when $\bar{\theta}_{fU}=
\theta_{fU}' + arg \: det \: M_{cd} =0 $. Thus, there will always exist CP
violating effects associated with the ultracolour groups. Moreover, because
the c--d mass matrix contains CP violating phases, and since the quark
mass matrices are proportional to the c--d and g--h mass matrices, strong
CP violation will be re--introduced when the vacuum enters the phase in
which condensates form. Consequently, the present model provides no
solution to the strong CP problem.

There are, however, three variants of the present model which would enable
us to solve the strong CP problem, the last being extremely speculative:

\noindent
{\em Variant 1:} \\
We have no real understanding of the c--d and g--h mass matrices --- we have
to assume that they descend from some other set of dynamics. We could
simply impose that the ultrastrong dynamics respects CP symmetry. Thus,
because the ultrastong dynamics respect CP, so would QCD.

\noindent
{\em Variant 2:} \\
We now have precisely the QCD situation over again: c and d transform under
$SU(3)_{fU}$, we have a mass matrix and ultrastrong CP violation. From
the absence of strong CP violation we know that there can be no
ultrastrong CP violation. Any dynamics giving c and d their mass may
be reparametrised, for convenience, by a Higgs sector. In particular,
therefore, we may have another PQ symmetry. Thus c and d acquire their
masses at some scale $\geq$ 1 TeV, and the new PQ symmetry relaxes
ultrastrong CP violation to zero. A new axion would exist, but would be
massive because $\Lambda_{fU} \geq$ 1 TeV (as opposed to $\Lambda_{QCD}
\sim$ 250 MeV), and, moreover, would not couple to quarks and leptons. We
would not expect to be aware of its existence until we probed the dynamics
of the c--d mass generation mechanism.

\noindent
{\em Variant 3:} \\
There is also an alternative solution which relies on the fact that
neutrinos are very light, if not massless. We know that, if the up quark
were massless, strong CP violation would not exist, since all CP violating
effects could be rotated onto the up quark mass term which, of course,
would be zero. If we could find a way of dumping all CP violation onto the
neutrino mass term, we would similarly have a solution. To include leptons
within the current formalism, however, presents some considerable difficulty
\cite{eight,seventeen}. The construction and viability of the quark
moose critically relies on the fact that quarks come in more than one
colour. Gauge anomaly cancellation in the flavour groups then requires
us to have an $SU(n_c)$ group for the ultracolour groups. This is a wonderful
coincidence, for we require that the ultracolour groups are non--abelian in
order that they may be asymptotically free. Unfortunately each flavour of
lepton comes in only one `colour.' Thus, we need strongly coupling $U(1)$
groups \cite{eight} !

In order to circumvent this difficulty, and in order to establish
quark--lepton symmetry, we suppose that leptons come in three colours
\cite{eighteen,seven}.
The $SU(3)$ of colour on the leptons is broken at some high scale
to $SU(2)$. We then allow the $SU(2)$ colour group to become confining,
so that the leptons transforming under the $SU(2)$ subgroup of the
original $SU(3)$ of colour on the leptons do not appear in the low
energy spectrum of particles. The remaining leptons, those transforming
under the broken generators of the $SU(3)$ colour group, are precisely
those which one sees at low energy. Henceforth we shall simply refer
to unbroken lepton colour, but with the understanding that there is a
mechanism for breaking lepton colour.
The lepton moose, then, is entirely equivalent to the quark moose of
figure 4. Notice that quark--lepton symmetry demands the existence of
right--handed neutrinos. A fourth generation neutrino would have to be
massive so as not to conflict with recent LEP data. Since the u,c,t (and T)
quark masses are parametrised as the mass matrix connecting the $c$ and $d$
particles, let the neutrino masses be parametrised as a mass matrix
connecting the $c'$ and $d'$ particles, where the $c'$ and $d'$ particles
are the lepton moose analogue of the quark moose $c$ and $d$ particles.

If we could establish a link between the quark and lepton moose, we could then
solve the strong CP problem. The simplest manner in which to
establish a connection is to demand that the ultracolour groups belonging
to the quark moose and the lepton moose be identical. Anomaly cancellation
tells us that the number of lepton colours equals the number of quark colours.
The $c'$--$d'$ mass matrix is tiny, if not identically zero, since the
neutrinos are very light. In the 4 generation model, only $(M_{c'd'})_{44}$
would be non--zero, so as to allow the fourth generation neutrino to become
massive. We may use our previous analysis to argue that all
$\theta$--parameters
can be moved onto $SU(3)_{fU}$, together with any phases in any of the mass
matrices. Now we exploit the massless neutrino by rotating the overall
$\bar{\theta}$ onto the (non--existent or small) $c'$--$d'$ mass terms. Hence
CP violating effects can be sucked away, and the massless neutrino solves the
CP problem. This conclusion is relatively model independent, needing only
a connection between the source of quark and lepton masses via non--abelian
groups.

\section{Conclusions}

Any dynamical theory of electroweak symmetry breaking and mass generation
based on gauged, chiral horizontal symmetries possesses a PQ symmetry which
is spontaneously broken at the weak scale. The appearence and breaking of
the PQ symmetry is, at first sight, a disaster, for one would assume that
the resulting pseudoscalar is the standard electroweak axion, which is ruled
out experimentally. We have seen, however, that the explicit breaking of
the PQ symmetry by flavour instantons provides a contribution to the axion
mass over and above the usual contribution from colour instantons. The
calculations performed above reveal that the axion mass is inversely
proportional to the scale characteristic of the new flavour and ultracolour
dynamics. Therefore, for large $\Lambda$, the flavour instanton contribution
to the axion mass is greatly suppressed, as decoupling would demand. {\it
This means that any dynamical theory based on principles similar to those
described above, and which requires a large $\Lambda$, is ruled out, due
to the non--existence of the standard axion.} However, all theories to
date suggest that $\Lambda$ must be large if the top quark mass is to be
consistent with constraints on the $\rho$--parameter. The simplest way to avoid
this conflict is to assume the existence of a fourth generation of quarks
and leptons. $\Lambda$ may then be $\sim$ 1 TeV, the T and B quarks give
no contribution to the $\rho$--parameter as they are mass degenerate, and,
moreover, the axion becomes sufficiently heavy to prevent it from being
a problem phenomenologically. These conclusions are relatively model
independent. Particular to our model, however, is the presence of a G.I.M.
mechanism which allows us to lower $\Lambda$ below $\sim$ $10^3$ TeV without
fear of phenomenological difficulties associated with flavour--changing
neutral--currents.

Unfortunately, the present model does not solve the strong CP problem: in
order to understand the absence of strong CP violation, we need to understand
the absence of ultrastrong CP violation. We have indicated some ways in
which such an understanding could be obtained. The most speculative extension
imposes a quark--lepton symmetry, and the masslessness of the neutrinos
supplies us with a natural understanding of the absence of ultrastong and
therefore strong CP violation.

\begin{center}
{\bf Acknowledgements}
\end{center}

S.F.K. is grateful to the S.E.R.C. for the support of an Advanced
Fellowship, and T.E. is grateful to the S.E.R.C. for the support of
a studentship.

\clearpage

\section*{Figure Captions}

\noindent
{\bf Figure 1} : The toy moose. The $SU(3)_c$ group is, of course, colour.

\noindent
{\bf Figure 2} : A `bubble' graph causing interactions between $t_L$ and
$t_R^c$, ultimately giving rise to a top quark condensate. The cross
represents a condensate insertion.

\noindent
{\bf Figure 3} : `Bubble' graphs allowing up and charm quarks to acquire
a mass from the top quark mass. The blob denotes a mass insertion.

\noindent
{\bf Figure 4} : Two moose bolted together, for four families.

\noindent
{\bf Figure 5} : The simple QCD moose.

\noindent
{\bf Figure 6} : A QCD effective instanton vertex explicitly breaking the
$U(1)_A$ quark symmetry.

\noindent
{\bf Figure 7} : An QCD instanton--induced, PQ violating contribution to the
axion potential.

\noindent
{\bf Figure 8} : The minimal instanton--induced, PQ violating contribution
to the axion potential in the moose model. $x$, $y$, and $z$ are the centres
of the instantons, and $\rho_x$, $\rho_y$, and $\rho_z$ are the instanton
sizes.

\end{document}